\documentclass[%
 reprint,
superscriptaddress,
 amsmath,amssymb,
 aps,
 pra,
prl,
floatfix,
]{revtex4-2}

\usepackage{graphicx}
\usepackage{dcolumn}
\usepackage{bm}
\usepackage{siunitx}
\usepackage{xcolor}
\usepackage{lmodern}
\usepackage[hidelinks]{hyperref}
\usepackage{orcidlink}
\usepackage{calc}

\usepackage{subcaption}

\usepackage{changes}
\definechangesauthor[name={Martijn}, color=magenta]{mlr}
\definechangesauthor[name={Julian}, color=orange]{jcb}
\definechangesauthor[name={Emily}, color=blue]{evk}

\begin{document}

\newcommand\ARCNL{Advanced Research Center for Nanolithography (ARCNL), Science Park 106, 1098 XG Amsterdam, The Netherlands.}

\newcommand\VU{Department of Physics and Astronomy, and LaserLaB, Vrije Universiteit Amsterdam, De Boelelaan 1100, 1081 HV Amsterdam, The Netherlands.}

\title{Beyond-EUV spectrum of highly-charged gadolinium}

\begin{abstract}
We report \emph{ab initio} relativistic calculations on the complex Gd$^{18+}$ and Gd$^{26+}$ highly-charged ions, which are relevant to beyond-extreme-ultraviolet (BEUV) light sources based on laser-produced gadolinium plasmas. Using the particle-hole configuration-interaction with many-body perturbation theory (CI+MBPT) method, we systematically saturate the $n=4$ configuration space and determine the influence of multiply excited states on the emissivity spectrum. For Gd$^{26+}$ we find behavior analogous to that of its isoelectronic counterpart Sn$^{12+}$, with strong contributions from multiply excited states, where even the quadruply excited states become significant at certain effective temperatures. In contrast, the industry-relevant 6.7\,nm emission of Pd-like Gd$^{18+}$ is dominated by the singly excited $4d^{10} {\ }^{1}S_{0}$ -- $4d^{9}4f {\ }^{1}P_{1}$ transition, while multiply excited states primarily contribute to out-of-band emission. The calculated spectra show good agreement with experiment and demonstrate that the importance of multiply excited states depends strongly on both charge state and plasma conditions.
\end{abstract}

\author{M.L. Reitsma\orcidlink{0000-0002-8255-7480}}
\affiliation{\ARCNL}
\author{J. Sheil\orcidlink{0000-0003-3393-9658}}
\affiliation{\ARCNL}
\affiliation{\VU}
\author{O.O. Versolato\orcidlink{0000-0003-3852-5227}}
\affiliation{\ARCNL}
\affiliation{\VU}

    \makeatletter
    \let\tmpaffiliation\affiliation
    \let\tmpauthor\author
    \let\tmpabstract\frontmatter@abstract@produce
    \let\frontmatter@abstract@produce\relax
    \let\frontmatter@finalspace\relax
    \maketitle
    
    \def\frontmatter@finalspace{\addvspace{18\p@}}
    \let\maketitle\frontmatter@maketitle
    \let\affiliation\tmpaffiliation
    \let\author\tmpauthor
    \let\frontmatter@abstract@produce\tmpabstract
    \let\frontmatter@title@produce\relax
    \makeatother

\author{E. V. Kahl\orcidlink{0000-0003-3923-7120}}
\affiliation{Pawsey Supercomputing Research Centre, 1 Bryce Avenue, Kensington, 6151, WA, Australia}
\author{J. C. Berengut\orcidlink{0000-0002-7366-1091}}
\affiliation{School of Physics, University of New South Wales, Sydney NSW 2052, Australia}

\date{\today}

\maketitle

\section{Introduction}

Laser-produced gadolinium plasmas are among the leading candidates for beyond-extreme-ultraviolet (BEUV) light sources emitting near 6.7\,nm~\cite{Otsuka2010}. This wavelength is of particular interest due to the high achievable reflectivity of La/B and related multilayer mirrors in this range~\cite{Makhotkin:13, kuznetsov2015high, liu2025physical}. Highly-charged Gd ions strongly emit in this region through $n=4-4$ transitions involving $4p$, $4d$, and $4f$ subshells~\cite{sugar1981ag, sugar1982resonance,higashiguchi2011extreme,suzuki2012observation,o2015sources}.

Similar to the case of 13.5\,nm emission from tin plasmas, predicting the amount of in-band radiation produced by Gd plasmas remains a challenging problem. Accurate atomic structure calculations are required for both interpreting experimental data and for use in collisional-radiative simulations employed in source design and optimization~\cite{li2011gd, kilbane2013,o2015spectroscopy,yin20166}.

One ion of particular importance is Pd-like Gd$^{18+}$. Plasma modeling and spectroscopic studies consistently identify this and neighboring charge states as major contributors to the emission near 6.7\,nm~\cite{li2011gd,Li2012, kambali2014spectroscopic,suzuki2015temperature}. The dominant resonance transition in this ion is the $4d^{10} {\ }^{1}S_{0}$ -- $4d^{9}4f {\ }^{1}P_{1}$ line~\cite{sugar1982resonance,higashiguchi2011extreme,suzuki2012observation}, but there are many multiply excited states within the $n=4$ shell that interact strongly with one another, and it remains unclear to what extent such states contribute to the spectrum. This situation is similar to that of highly-charged tin ions, where strong configuration mixing between $n=4$ configurations was shown to give rise to complex spectra dominated by multiply excited states~\cite{darcy2009transitions,torretti2020prominent}, and where our recent study showed that a complete treatment of configuration interaction within the $n=4$ shell is essential for obtaining accurate spectra in such systems~\cite{Reitsma2026arxiv}. It is an open question whether BEUV emission from highly-charged Gd ions is affected by similar multiply excited states and configuration-mixing effects.

In this work, we apply the particle-hole configuration-interaction with many-body perturbation theory (CI+MBPT) method to investigate the electronic structure and spectra of Gd$^{18+}$ and Gd$^{26+}$ originating from $n=4-4$ transitions. The former is one of the most important charge states contributing to BEUV emission near 6.7\,nm, and the latter is isoelectronic with the Sn$^{12+}$ ion, providing a useful point of comparison to our previous study. We systematically analyse the role of singly and multiply excited configurations, quantify the convergence of the spectrum with respect to $n=4$ configuration space saturation, and determine the relative contributions of these configurations to the spectrum at effective temperatures corresponding to different plasma conditions. We show that the in-band emission of Gd$^{18+}$ is dominated by the singly excited $4d^{10} {\ }^{1}S_{0}$ -- $4d^{9}4f {\ }^{1}P_{1}$ transition, whereas multiply excited states play a minor role and are mainly relevant for out-of-band emission at higher temperatures. For Gd$^{26+}$, in contrast, multiply excited states are significantly more important at higher temperatures, where the spectrum even contains contributions from quadruply excited states.

\section{Method}
\subsection{CI+MBPT}
We used the particle-hole CI+MBPT (configuration interaction combined with many-body perturbation theory) method~\cite{berengut16pra} implemented in the AMBiT program~\cite{Kahl2019}, which extends the CI+MBPT method~\cite{Dzuba1996} to allow for valence holes to be treated in CI. 
The Dirac-Coulomb Hamiltonian for our calculations is (in atomic units $\hbar = m_e = e = 1$)
\begin{equation}
    H_{DC} = \sum_i h_D(i) + \sum_{i<j} 1/r_{ij},
\end{equation}
where $h_D$ is the one-electron Dirac Hamiltonian,
\begin{equation}
    h_{D}(i)=c\, \boldsymbol \alpha_{i}\cdot \mathbf{p}_{i}+c^{2}(\beta _{i}-1)+V_\text{nuc}(i).
\end{equation}
Here, $\boldsymbol \alpha$ and $\beta$ are the four-dimensional Dirac matrices, and a Fermi charge distribution~\cite{Visscher97} is used for the nuclear potential $V_\text{nuc}$. Additionally, we calculated the magnitude of the frequency-independent Breit interaction, self-energy, and (Uehling) vacuum polarization effects using an effective potential~\cite{flambaum2005radiative,ginges2016atomic,ginges2016qed} in separate calculations.

We begin with a self-consistent Dirac-Hartree-Fock (DHF) calculation. In the Gd$^{18+}$ (Gd$^{26+}$) case, we treated $N=46$ ($N=38$) electrons in the V$^{N}$ approximation corresponding to the configuration-averaged ground state [Kr]~$4d^{10}$ ([Kr]~$4d^2$). We then generate our orbital basis by diagonalizing a complete set of B-splines over the DHF potential. The $4p$, $4d$, and $4f$ shells are treated as valence and correlated in our CI expansion as explained below.

In the CI+MBPT method, core-valence excitations are treated at second order in the residual Coulomb interaction by modifying the one and two-particle integrals in the CI procedure~\cite{Dzuba1996}. Effective three-body operators are separately added to the Hamiltonian matrix elements. The core-valence diagrams include virtual orbitals up to 30\textit{spdfghi} (i.e., $n\leq30$ and $l\leq6$). We also performed small-scale tests up to $n=35$ and $l=7$ to confirm that the MBPT contributions are well converged at this level. Using AMBiT's implementation of particle-hole CI+MBPT theory, we place the Fermi level above the $4p$ shell and treat $4p$ vacancies as holes, which reduces the size of subtraction diagrams compared to treating $4p$ electrons as valence particles. 

The CI space is built following the same approach and naming conventions as in Ref.~\cite{Reitsma2026arxiv}. Configurations are added in sequence according to the number of excitations from the ground state, designated as the m-fold, and are limited to excitations within the $n=4$ shell. The general energy level structure arising from these configurations is seen in Fig.~\ref{fig:Gd_grotrian_both}, where we can also see that neighboring m-fold states are separated by roughly 180\,eV for the two ions under consideration. Transitions between m-fold and k-fold states are designated by m$\rightarrow$k. A CI space that contains all configurations from the 0-fold up to and including the m-fold set is written as \{m\}.

The CI space of Gd$^{26+}$ is built in the same way as its isoelectronic partner Sn$^{12+}$ in our previous study, which is summarized in Table~\ref{tab:Gd26+_configs}. The difference in this case is the ionization potential (IP), which is much higher relative to the electronic states than for Sn$^{12+}$, such that the 4-fold excited states are now below the IP. These states are thus not autoionizing and contribute significantly to the emissivity. We therefore also need to include the 4$\rightarrow$3 transitions between the 4-fold and 3-fold states. The element gadolinium is also heavier than tin, meaning that relativistic effects are expected to be more pronounced.
\begin{table}[hbtp]
    \centering
    \begin{tabular}{ll}
    \hline
    \hline
        Set & Configurations \\
    \hline
        0-fold & $4p^6 4d^2$ \\
        1-fold & $4p^5 4d^3$, $4p^6 4d^1 4f^1$\\
        2-fold & $4p^4 4d^4$, $4p^6 4f^2$, $4p^5 4d^2 4f^1$ \\
        3-fold & $4p^3 4d^5$, $4p^5 4d^1 4f^2$, $4p^4 4d^3 4f^1$ \\
        4-fold & $4p^2 4d^6$, $4p^3 4d^4 4f^1$, $4p^4 4d^2 4f^2$, $4p^5 4f^3$ \\
        5-fold & $4p^1 4d^7$, $4p^2 4d^5 4f^1$, $4p^3 4d^3 4f^2$, $4p^4 4d^1 4f^3$ \\
        6-fold & $4p^0 4d^8$, $4p^1 4d^6 4f^1$, $4p^2 4d^4 4f^2$, $4p^3 4d^2 4f^3$, $4p^4 4f^4$ \\
    \hline
    \hline
    \end{tabular}
    \caption{The $n=4$ configurations included in each m-fold configuration set for Gd$^{26+}$.}
    \label{tab:Gd26+_configs}
\end{table}

For Gd$^{18+}$, the CI space is built starting from the ground state [Kr]$4d^{10}$, from which we generate excited configurations starting within the $n=4$ shell up to 5-fold excitations, which are listed in Table~\ref{tab:Gd18+_configs}.
\begin{table}[hbtp]
    \centering
    \begin{tabular}{ll}
    \hline
    \hline
        Set & Configurations \\
    \hline
        0-fold & $4p^6 4d^{10}$ \\
        1-fold & $4p^6 4d^9 4f^1$ \\
        2-fold & $4p^6 4d^8 4f^2$, $4p^5 4d^{10} 4f^1$ \\
        3-fold & $4p^6 4d^7 4f^3$, $4p^5 4d^9 4f^2$ \\
        4-fold & $4p^6 4d^6 4f^4$, $4p^5 4d^8 4f^3$, $4p^4 4d^{10} 4f^2$ \\
        5-fold & $4p^6 4d^5 4f^5$, $4p^5 4d^7 4f^4$, $4p^4 4d^9 4f^3$ \\
    \hline
    \hline
    \end{tabular}
    \caption{The $n=4$ configurations included in each m-fold configuration set for Gd$^{18+}$.}
    \label{tab:Gd18+_configs}
\end{table}

\subsection{Effective temperatures}
To analyze the contribution of each ion to the emissivity of the plasma, we determined the Boltzmann-weighted emissivity of a single ion by multiplying each transition rate $gA$ by the Boltzmann population factor $\exp(-E_i^\text{upper}/kT_\text{eff})$. In this case, we use an effective temperature $T_\text{eff}$ instead of the plasma's electron temperature $T_\text{e}$. Sheil et al.~\cite{Sheil2026} have shown that the atomic populations of non-local thermodynamic equilibrium plasmas can often be described by a Boltzmann distribution at effective temperatures different from $T_\text{e}$. The effective temperature is derived as
\begin{equation}
\frac{1}{T_\text{eff}} = \frac{1}{T_\text{e}} + \frac{1}{E_{21}}\ln\left[1 + \frac{C E_{21}^3 T_\text{e}^{1/2}}{n_\text{e}\bar{g}}\right],
\label{eq:Teff}
\end{equation}
where $C=2.69 \times 10^{12}$\,eV$^{-7/2}$\,cm$^{-3}$, $n_\text{e}$ is the electron density, the Gaunt factor $\bar{g} = 0.6 + 0.28\ln[1+(0.562+1.4\mathcal{F})/(\mathcal{F}+1.4\mathcal{F}^2)]$, $\mathcal{F} \equiv E_{21}/T_\text{e}$~\cite{mewe1972interpolation,shevelko1997atoms,HANSEN2006272} (appropriate for $\Delta n=0$ transitions), and $E_{21}$ is the level energy separation in the simplified two-level model that was used to derive this equation.
In the current study, $E_{21}$ is taken to be $\approx 180\,\text{eV}$, which is the energy separation of the 0-fold and 1-fold states. 
The electron densities that are considered in this study are $10^{19}$, $10^{20}$, and $10^{21}$\,cm$^{-3}$, as these cover the range of critical densities for lasers with wavelengths ranging from 10.6 to 1.064\,$\mu$m. 
Values for $T_\text{eff}$ are determined at the different densities for each ion by using (\ref{eq:Teff}), which depends on $T_\text{e}$. Values for $T_\text{e}$ are obtained from charge state distribution calculations of Ref.~\cite{chung2005flychk}. Here, $T_\text{e}$ is taken as the optimal temperature at which the charge state distribution is at a maximum for the ion and density in question. The resulting values of $T_\text{eff}$ are shown in Table~\ref{tab:ne_Te_Teff}, which were used to produce Boltzmann-weighted emissivity spectra for both ions. Note that $T_\text{e}$ decreases with increasing critical density, which is a consequence of collisional-radiative balance; at higher densities, collisional processes become more efficient and shift the charge-state distribution towards lower ionization species, so a lower electron temperature is required to maximize a given charge state.
\begin{table}[hbtp]
    \centering
    \begin{tabular}{c c c c}
    \hline
    \hline
        Ion & $n_\text{e}$ (cm$^{-3}$) & $T_\text{e}$ (eV) & $T_\text{eff}$ (eV) \\
    \hline
        Gd$^{18+}$ & $10^{19}$ & 117 & 38 \\
                   & $10^{20}$ & 74 & 51 \\
                   & $10^{21}$ & 68 & 64 \\
    \noalign{\vspace{4pt}}
        Gd$^{26+}$ & $10^{19}$ & 445 & 44 \\
                   & $10^{20}$ & 335 & 89 \\
                   & $10^{21}$ & 196 & 154 \\
    \hline
    \hline
    \end{tabular}
    \caption{Effective temperatures $T_\text{eff}$ for each gadolinium ion corresponding to different plasma conditions with electron density $n_\text{e}$ and temperature $T_\text{e}$.}
    \label{tab:ne_Te_Teff}
\end{table}

\section{Results and discussion}
We first examine the level structure and convergence of the calculated $n=4-4$ transition energies with respect to CI space saturation, as well as the influence of core-valence correlation, Breit, and QED corrections. We then investigate the resulting emissivity spectra at effective temperatures relevant to laser-produced gadolinium plasmas, where we determine the relative importance of singly and multiply excited states. We compare Gd$^{18+}$, with its technologically-relevant 6.7\,nm emission, to Gd$^{26+}$, which is of interest through its connection with the isoelectronic Sn$^{12+}$ ion.

\subsection{Structure and transitions}

\begin{figure}
    \centering
    \begin{subfigure}{0.4\linewidth}
        \includegraphics[width=\textwidth]{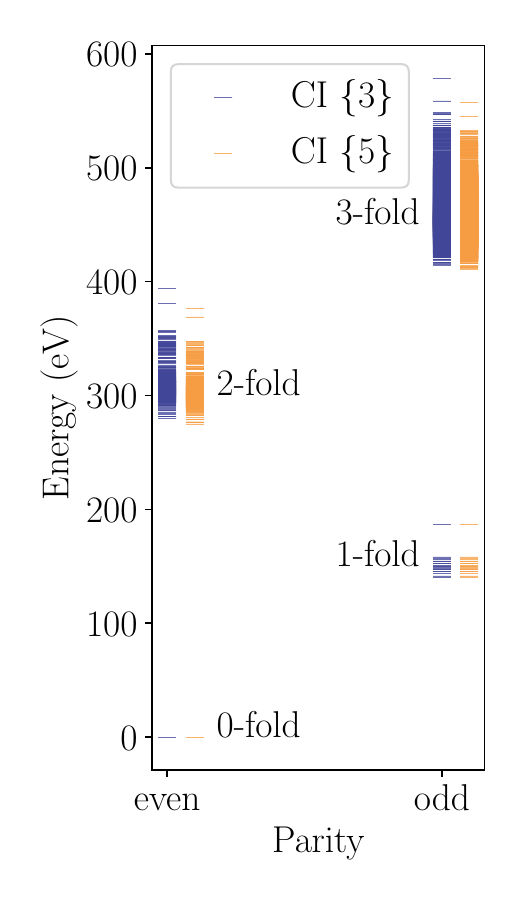}
    \end{subfigure}
    \begin{subfigure}{0.4\linewidth}
        \includegraphics[width=\textwidth]{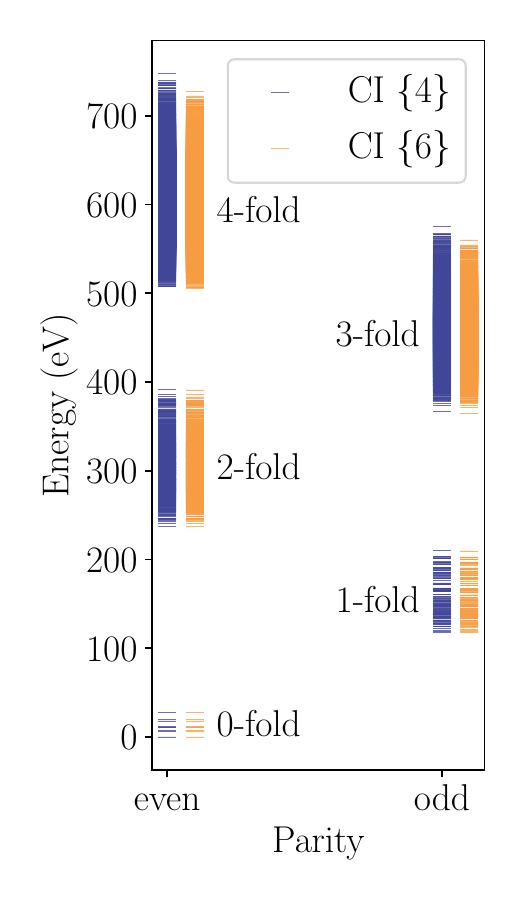}
    \end{subfigure}
    \caption{Level diagrams of Gd$^{18+}$ (left) and Gd$^{26+}$ (right) calculated for a small and large CI space, showing the effect of configuration mixing on the energies.}
    \label{fig:Gd_grotrian_both}
\end{figure}

Figure~\ref{fig:Gd_grotrian_both} shows the calculated $n=4$ level structure for both ions at increasing CI saturation. In both cases, the neighboring m-fold states are separated by approximately 180\,eV. We note that only the higher-energy (multiply excited) states are visibly affected by the additional correlation, when comparing the smaller CI space (blue) to the larger CI space (yellow). The main effect is a downward shift in energy for those multiply excited states, which is caused by configuration mixing with even higher-lying excited states (which are outside the range of the plots). Specifically, the 1-fold energies are shifted by a negligible 0.08\,eV relative to the ground state on average for Gd$^{18+}$. The average shifts of the 2-fold and 3-fold states are significant, as their energies are decreased by 5.5 and 5.0\,eV respectively, owing to the strong configuration mixing with the 4- and 5-fold configurations. A similar behavior occurs for Gd$^{26+}$, but in this case the higher ionization potential places the 4-fold excited states below the continuum, thus making radiative 4$\rightarrow$3 transitions possible. Configuration mixing has a significant impact on the calculated transition wavelengths and emissivity spectra that are discussed next, through both these energy-level shifts and a redistribution of oscillator strengths.

We first consider the calculated transition energies of Gd$^{26+}$. The resulting mean wavelengths of the $n=4-4$ transitions in Gd$^{26+}$ can be found in Table~\ref{tab:Gd26+_mean_nm}. The weighted mean $\mu$ of a transition set is computed by weighting each transition energy $\Delta E_i = E_i^\text{upper} - E_i^\text{lower}$ by its degeneracy-weighted transition probability $g_iA_i$ as 
\begin{equation}
    \mu_E = \frac{\sum_i g_iA_i \Delta E_i}{\sum_i g_iA_i}\ \text{or}\ \mu_\lambda = \frac{hc}{\mu_E}.
\end{equation}
\begin{table}[hbtp]
    \centering
    \begin{tabular}{ccccc}
    \hline
    \hline
    CI space & 1$\rightarrow$0 & 2$\rightarrow$1 & 3$\rightarrow$2 & 4$\rightarrow$3 \\
    \hline
    \{1\} & 6.95 & -- & -- & -- \\
    \{2\} & 6.72 & 7.04 & -- & -- \\
    \{3\} & 7.04 & 6.82 & 7.14 & -- \\
    \{4\} & 7.01 & 7.12 & 6.92 & 7.29 \\
    \{5\} & 7.04 & 7.10 & 7.19 & 7.07 \\
    \{6\} & 7.04 & 7.11 & 7.18 & 7.32 \\
    
    \hline
    $\Delta$MBPT & 0.10 & 0.14 & 0.17 & 0.14 \\
    $\Delta$Breit & 0.02 & 0.02 & 0.02 & 0.02 \\
    $\Delta$QED      & 0.002 & 0.001 & 0.001 & 0.003 \\
    \hline
    
    Total & 7.17 & 7.28 & 7.37 & 7.49 \\

    \hline
    \hline
    \end{tabular}
    \caption{Mean wavelength (nm) of the Gd$^{26+}$ $n=4-4$ spectrum, for different CI spaces and for single (1$\rightarrow$0) up to quadruple (4$\rightarrow$3) transitions.}
    \label{tab:Gd26+_mean_nm}
\end{table}

\begin{figure}[htbp]
    \centering
    \includegraphics[width=1.0\linewidth]{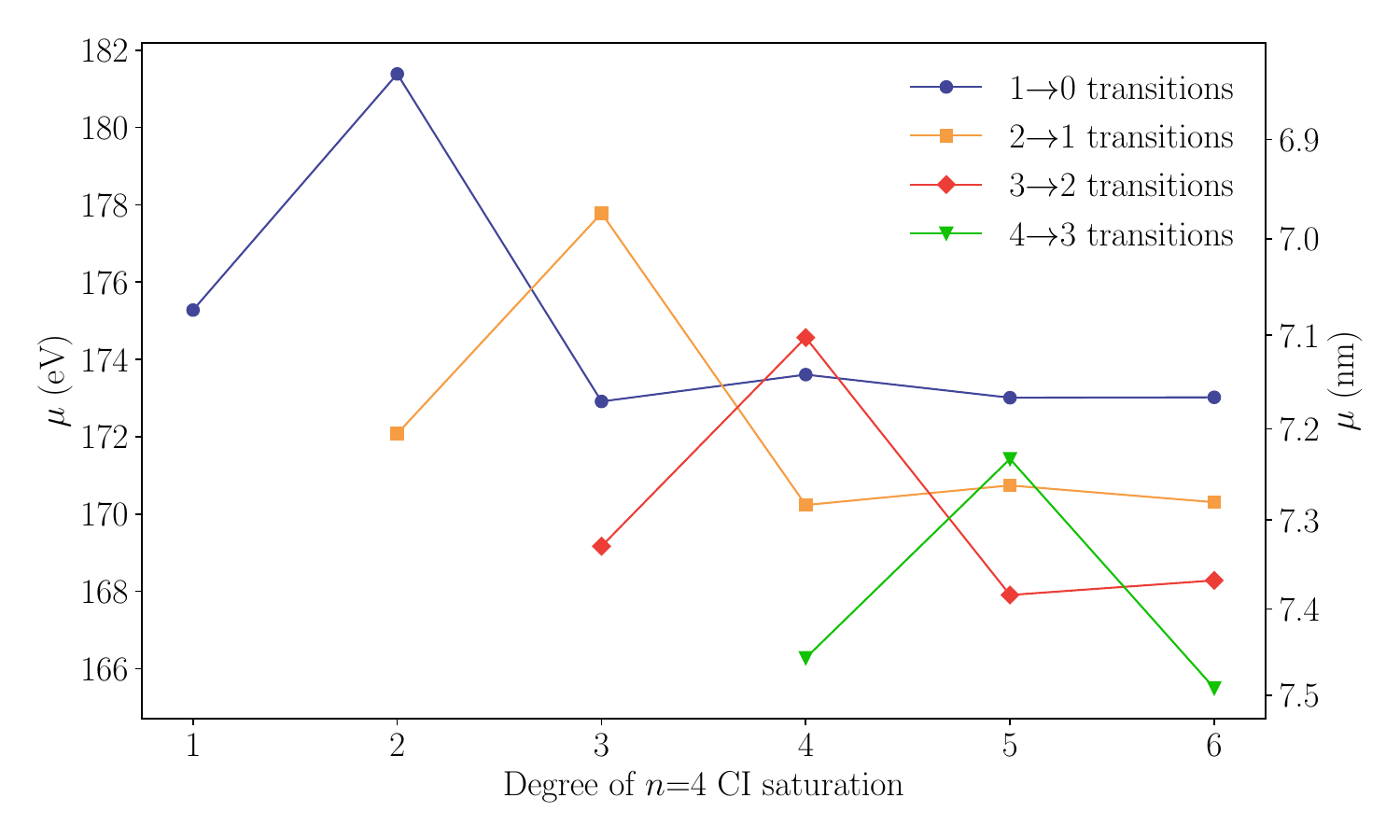}
    \caption{Mean wavelength of the Gd$^{26+}$ $n=4-4$ spectrum as a function of the CI space shown separately for each transition class.}
    \label{fig:Gd26+_alps_absolute}
\end{figure}

These results are also plotted in Fig.~\ref{fig:Gd26+_alps_absolute}, where the core-valence, Breit and QED corrections are already included. The convergence behavior with respect to the CI space is very similar to the Sn$^{12+}$ case of Ref.~\cite{Reitsma2026arxiv} for the 1$\rightarrow$0, 2$\rightarrow$1 and 3$\rightarrow$2 transitions, where the positions of the spectra are well-converged at the CI \{5\} level. The 4$\rightarrow$3 transitions, however, are relevant in this case, because of the much higher IP of the Gd$^{26+}$ ion. The wavelength of these transitions is shifted by a significant 0.25\,nm going from CI space \{5\} to \{6\}. This is a crucial detail; where in the tin case it was shown that the \{5\} CI space is sufficiently converged, this is not the case for the 4-fold excited states of Gd$^{26+}$.

In contrast, Gd$^{18+}$ has mean transition wavelengths close to the narrow 0.04\,nm reflectivity bandwidth centered at 6.76\,nm. The calculated mean wavelengths for the different transition sets are found in Table~\ref{tab:Gd18+_mean_nm}. Looking at the effect of CI saturation between the \{3\} and \{5\} spaces, there is a difference in wavelength of about 0.3\,nm for the 2$\rightarrow$1 transitions, and 0.1\,nm for the 3$\rightarrow$2 transitions, which corresponds to an energy difference of 8.8 and 2.2\,eV respectively. Notably, this is considerably larger than what is expected from the energy level shifts only, since the strong configuration mixing also causes a redistribution of oscillator strengths.

It is noteworthy that, though they are still smaller than core-valence corrections, the Breit and QED corrections are much more important than they were for the tin case; the energy change due to the Breit interaction is over three times larger, while it is over ten times larger for the QED correction. Moreover, the size of the Breit correction of 0.02\,nm is of the same order as the reflectivity bandwidth of La/B mirrors of 0.04\,nm at 6.76\,nm. 
\begin{table}[hbtp]
    \centering
    \begin{tabular}{cccc}
    \hline
    \hline
    CI space & 1$\rightarrow$0 & 2$\rightarrow$1 & 3$\rightarrow$2\\
    \hline
    \{1\} & 6.53 & -- & -- \\
    \{2\} & 6.28 & 6.64 & -- \\
    \{3\} & 6.64 & 6.41 & 6.74 \\
    \{4\} & 6.62 & 6.74 & 6.53 \\
    \{5\} & 6.65 & 6.72 & 6.82 \\
    
    \hline
    $\Delta$MBPT    & 0.08 & 0.09 & 0.09 \\ 
    $\Delta$Breit   & 0.01 & 0.01 & 0.01 \\ 
    $\Delta$QED     & -0.0004 & -0.0003 & -0.0002 \\ 
    \hline
    
    Total & 6.74 & 6.82 & 6.93 \\
    \hline
    \hline
    \end{tabular}
    \caption{Mean wavelength of the Gd$^{18+}$ $n=4-4$ spectrum, for different CI spaces and for single (1$\rightarrow$0) up to triple (3$\rightarrow$2) transitions.}
    \label{tab:Gd18+_mean_nm}
\end{table}

\begin{figure}[htbp]
    \centering
    \includegraphics[width=1.0\linewidth]{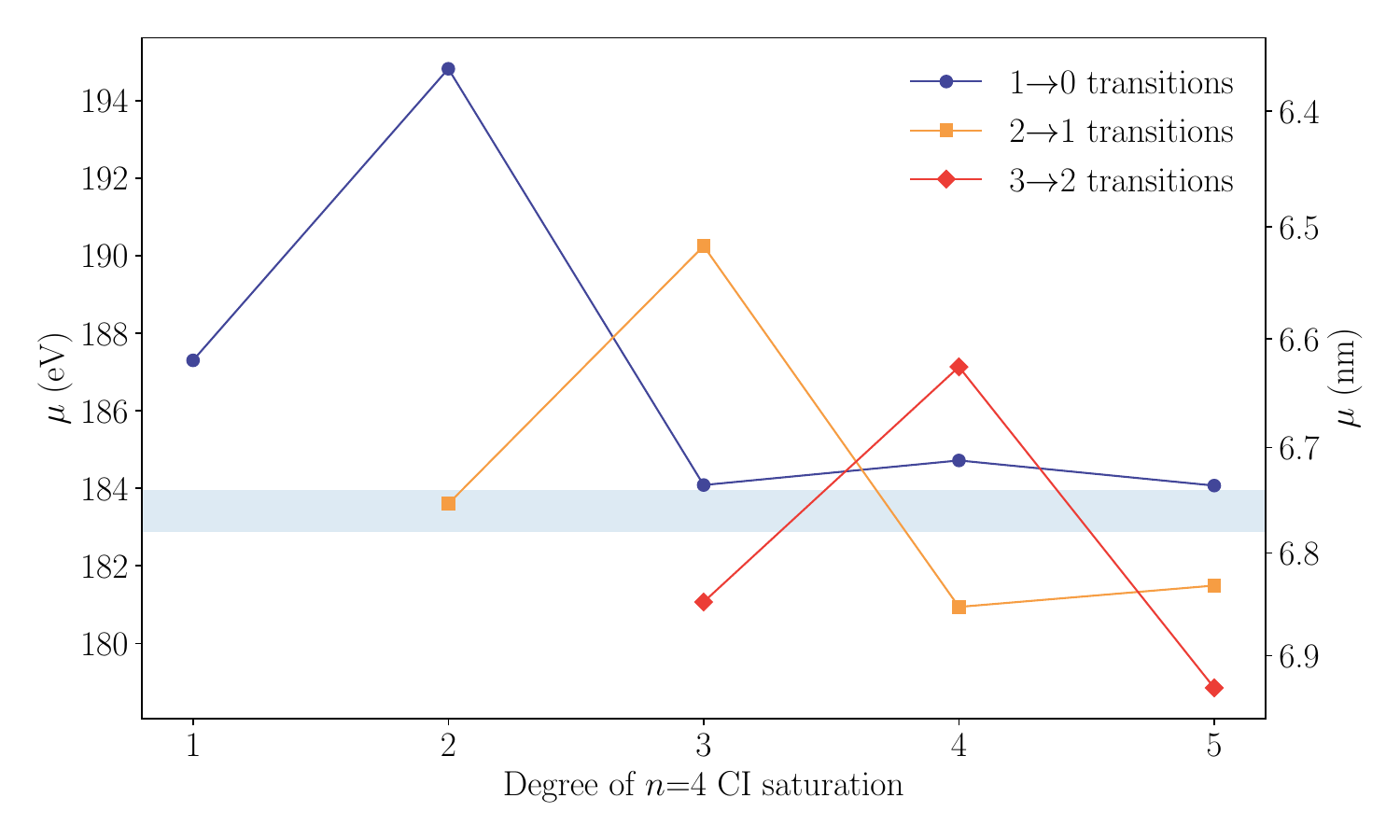}
    \caption{Mean wavelength of the Gd$^{18+}$ $n=4-4$ spectrum as a function of the CI space shown separately for each transition class. The shaded region shows a 0.7\% reflectivity bandwidth of La/B mirrors around 6.7\,nm.}
    \label{fig:Gd18+_alps_absolute}
\end{figure}

We see in Fig.~\ref{fig:Gd18+_alps_absolute}, which shows the mean wavelength as a function of $n=4$ CI saturation including corrections, that the convergence pattern of Gd$^{18+}$ is almost identical to that of Gd$^{26+}$. We note that the mean wavelength for 1$\rightarrow$0 is closest to 6.7\,nm, which is dominated by a single strong $4d^{10} {\ }^{1}S_{0}$ -- $4d^{9}4f {\ }^{1}P_{1}$ line. The mean wavelength of the multiply excited transitions is slightly longer close to 6.8\,nm and 6.9\,nm. 

The $4d^{10} {\ }^{1}S_{0}$ -- $4d^{9}4f {\ }^{1}P_{1}$ line of Gd$^{18+}$ may be compared to HULLAC calculations of Sasaki et al.~\cite{Sasaki2022}, where a wavelength of 6.6\,nm is reported. Cowan-code calculations by Li et al.~\cite{li2011gd} predict a peak near 6.76\,nm. High-voltage spark measurements by Sugar and Kaufman~\cite{sugar1982resonance} place the line at 6.7636\,nm. Our calculated value of 6.74\,nm lies within 0.02\,nm of the experimental result. The remaining disagreement can be attributed to correlation with configurations that were not included in this study, mainly the current limitation to the \{5\} CI space that is due to a technical difficulty with obtaining the \{6\} results. Correlation with $n>4$ shells may also have an effect and requires further study.

\subsection{Boltzmann-weighted emissivity}

All emissivity spectra presented in this work are convolved with a Gaussian profile of 0.2\,eV width to simulate experimental resolution~\cite{Kume2024}.

\begin{figure}[htbp]
    \centering
    \includegraphics[width=1.0\linewidth]{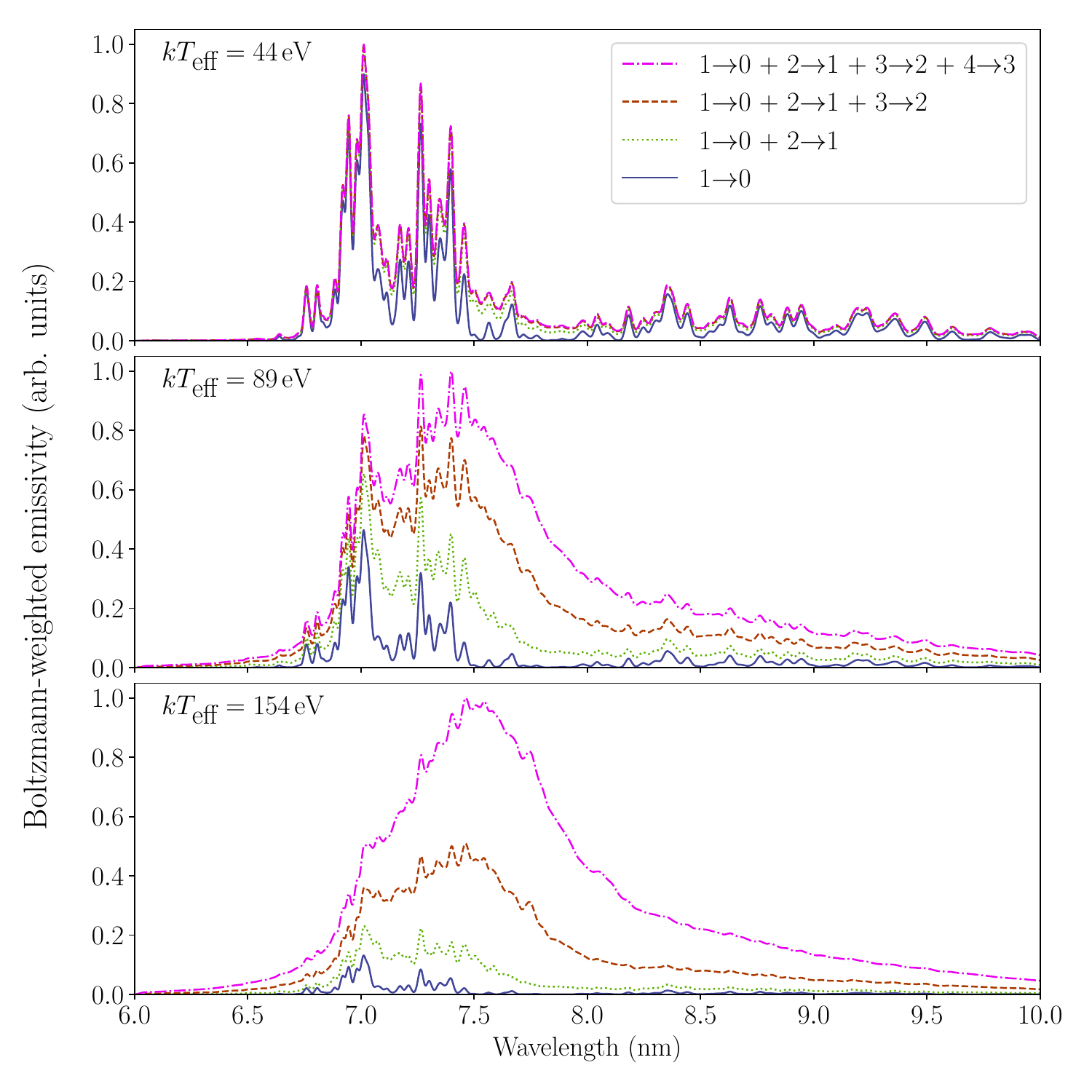}
    \caption{Boltzmann-weighted emissivity $gA\exp(-E_i^\text{upper}/kT_\text{eff})$ of $n=4-4$ transitions of Gd$^{26+}$ at three distinct effective temperatures. The calculations were performed using the \{6\} CI space and include $\Delta$MBPT, $\Delta$Breit, and $\Delta$QED corrections.}
    \label{fig:Gd26+_multiple}
\end{figure}

The Boltzmann-weighted emissivities for $n=4-4$ transitions of Gd$^{26+}$ are shown in Fig.~\ref{fig:Gd26+_multiple} for the different effective temperatures. At $T_\text{eff}=44$\,eV, corresponding to $n_\text{e}=10^{19}$\,cm$^{-3}$, the spectrum is determined solely by the 1$\rightarrow$0 transitions, with barely any contribution from multiply excited states. The optimal effective temperature at higher densities is significantly higher, and this results in the emergence of large contributions from multiply excited states at 89\,eV. At an effective temperature of 154\,eV, the 4$\rightarrow$3 transitions from quadruply excited states are the dominant source of emission. The change from a spectrum dominated by 1-fold excited states at low $T_\text{eff}$ to one dominated by 4-fold excited states at high $T_\text{eff}$ illustrates that the importance of many-electron excitations is strongly temperature dependent, which is a crucial consideration for the emissivity of dense plasma sources in particular.

\begin{figure}[htbp]
    \centering
    \includegraphics[width=1.0\linewidth]{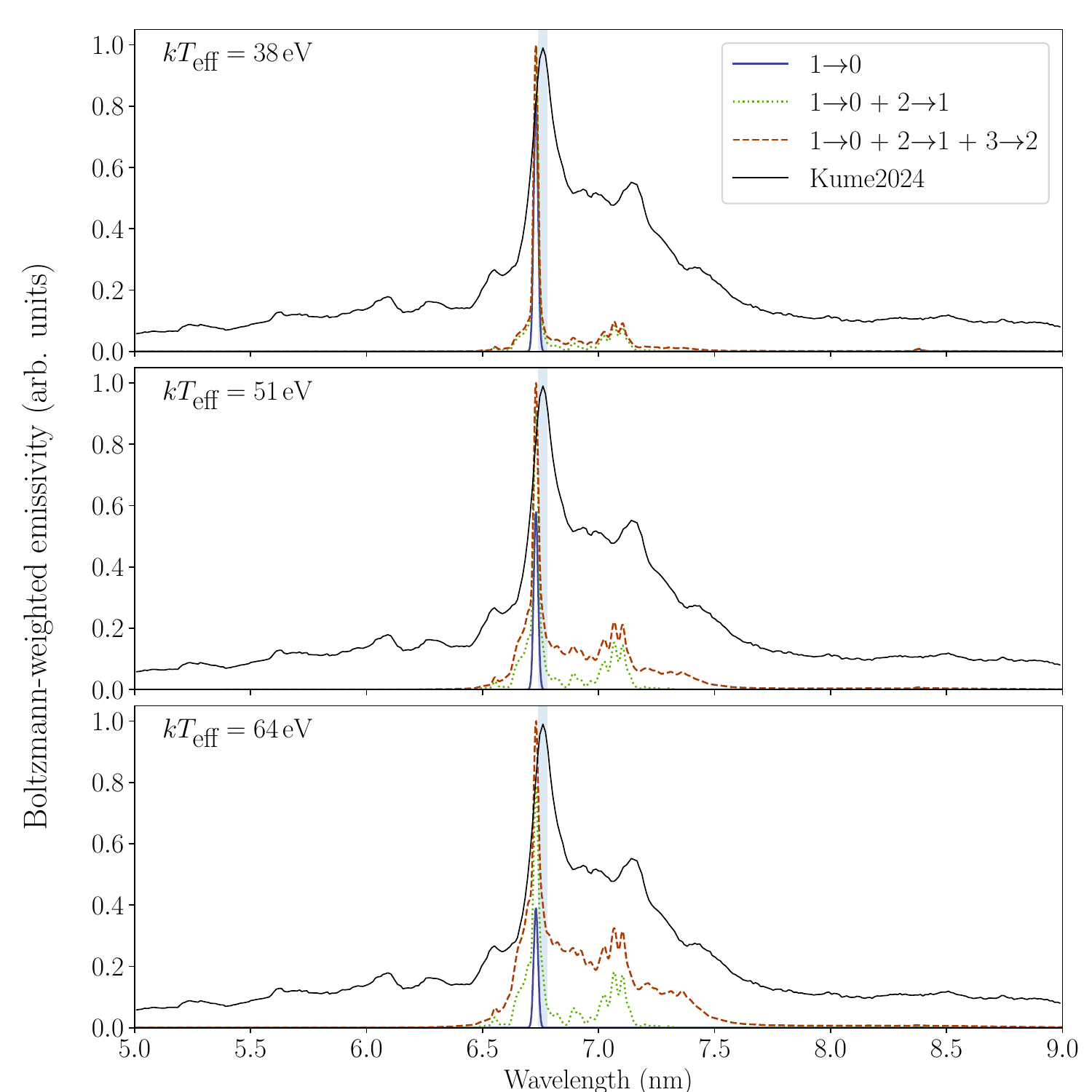}
    \caption{Boltzmann-weighted emissivity $gA\exp(-E_i^\text{upper}/kT_\text{eff})$ of $n=4-4$ transitions of Gd$^{18+}$ at three distinct effective temperatures. The shaded region shows a 0.6\% bandwidth around 6.76\,nm, corresponding to the reflective region of La/B mirrors. The calculations were performed using the \{5\} CI space  and include $\Delta$MBPT, $\Delta$Breit, and $\Delta$QED corrections. The solid black line shows the experimental spectrum of a gadolinium plasma measured by Kume et al.~\cite{Kume2024}.}
    \label{fig:Gd18+_multiple}
\end{figure}

The situation is different for Gd$^{18+}$, which would not require such high temperatures to be produced at these densities, as is shown in Fig.~\ref{fig:Gd18+_multiple}. At an effective temperature of 38\,eV, the singly excited $4d^{10} {\ }^{1}S_{0}$ -- $4d^{9}4f {\ }^{1}P_{1}$ transition line is the only major contribution to the spectrum, which is also the case at 51\,eV, though at both 51\,eV and 64\,eV the populations of the 2-fold and 3-fold excited states increase and transitions from these states become important. This shows that multiply excited states for Gd$^{18+}$ are crucial to describe the full spectrum, especially at higher temperatures, corresponding to higher-density plasmas. Their contribution to the spectrum near the industry-relevant 6.7\,nm wavelength, however, is limited for this ion, although their inclusion is important for the accuracy of the calculated line positions.
In the same figure, a comparison is made to an experimental spectrum of a laser-produced gadolinium plasma~\cite{Kume2024}. 
Even though our calculated spectrum only includes a single ion, while the experiment contains the spectrum of several charge states, the agreement between theory and experiment of the peak positions is remarkable. The lower peaks to the right of the main peak also appear in the experimental spectrum. In the theoretical spectrum, those peaks originate from 2$\rightarrow$1 transitions, which may indicate that transitions between multiply excited states contribute to the experimentally observed spectrum.

\section{Conclusion}
We have performed ab initio relativistic CI+MBPT calculations of the electronic structure and $n=4-4$ transitions of the highly-charged Gd$^{18+}$ and Gd$^{26+}$ ions, which are relevant to laser-produced gadolinium plasmas for next-generation BEUV lithography. By systematically expanding the $n=4$ configuration space, we demonstrate the influence of singly and multiply excited configurations on the predicted transition energies, and quantify the relative contributions of singly and multiply excited states to the emissivity at relevant plasma temperatures, using the effective temperature model.

For Gd$^{26+}$ we find that the convergence behavior of the transition energies with increasing $n=4$ CI saturation is analogous to that of its isoelectronic counterpart Sn$^{12+}$. At a low effective temperature corresponding to a low-density ($n_\text{e}=10^{19}$cm$^{-3}$), the emission is dominated by 1$\rightarrow$0 transitions, while transitions from multiply excited states only start contributing at higher $T_\text{eff}$. The higher ionization potential of Gd$^{26+}$ places quadruply excited states below the continuum. As a result, 4$\rightarrow$3 transitions become the dominant source of emission at sufficiently high temperatures, and therefore cannot be neglected in spectral modeling when considering such plasma conditions.

The case of Gd$^{18+}$ is substantially different. Although configuration interaction with multiply excited states substantially modifies the level structure and contributes to the out-of-band emission, the industry-relevant 6.7\,nm emission remains dominated by the singly excited $4d^{10} {\ }^{1}S_{0}$ -- $4d^{9}4f {\ }^{1}P_{1}$ transition. Contributions from 2$\rightarrow$1 and 3$\rightarrow$2 transitions become increasingly important at higher temperatures, but they appear mostly outside of the narrow 6.7\,nm wavelength band. Remarkable agreement with experiment is observed, and may indicate a direct observation of transitions between multiply excited states.

The shifts due to the inclusion of core correlation are comparable to the acceptance bandwidth of typical multilayer mirrors, demonstrating that accurate prediction of the BEUV spectrum requires an explicit treatment of both core-valence correlation and a saturated configuration interaction of the valence space.

The present calculations establish a clear description of the $n=4$ electronic structure of Gd$^{18+}$ and Gd$^{26+}$, demonstrating how the importance of multiply excited states strongly depends on the charge state in question and the plasma conditions that are considered.

\section{Acknowledgments}
This work was conducted at the Advanced Research Center for Nanolithography (ARCNL), a public-private partnership between the University of Amsterdam (UvA), Vrije Universiteit Amsterdam (VU), Rijksuniversiteit Groningen (UG), the Dutch Research Council (NWO), and the semiconductor equipment manufacturer ASML. 
This work used the Dutch national e-infrastructure with the support of the SURF Cooperative using grant no. EINF-13731. 
This publication is part of the project ARIES with file number 20152 of the research programme VENI which is financed by NWO. This work was funded by ERC CoG MOORELIGHT 101086839. 
EVK's work was supported by resources provided by the Pawsey Supercomputing Research Centre’s Setonix Supercomputer~\cite{Setonix2023} and PULSE research software engineering collaboration, with funding from the Australian Government and the Government of Western Australia.

\bibliography{references}

\end{document}